\documentclass[twocolumn,showpacs,preprintnumbers,amsmath,amssymb]{revtex4}

\usepackage{graphicx}
\usepackage{dcolumn}
\usepackage{bm}

\begin{document}
\title{Quantized Anomalous Hall Effect in Two-Dimensional Ferromagnets
\\
- Quantum  Hall Effect from Metal -}
%
\author{Masaru Onoda$^1$}
\email{m.onoda@aist.go.jp}
\author{Naoto Nagaosa$^{1,2}$}
\email{nagaosa@appi.t.u-tokyo.ac.jp}
\affiliation{
$^1$Correlated Electron Research Center (CERC),
National Institute of Advanced Industrial Science and Technology (AIST),
Tsukuba Central 4, Tsukuba 305-8562, Japan\\
$^2$Department of Applied Physics, University of Tokyo,
Bunkyo-ku, Tokyo 113-8656, Japan
}
%
%
\date{\today}
%
\begin{abstract}
We study the effect of disorder on the anomalous Hall effect (AHE)
in two-dimensional ferromagnets. The topological nature
of AHE leads to the integer quantum Hall effect from a
metal, i.e., the quantization of $\sigma_{xy}$ induced by the
localization except for the few extended states carrying
Chern number. Extensive numerical study on a model reveals that
Pruisken's two-parameter scaling theory holds even when
the system has no gap with the overlapping multibands
and without the uniform magnetic field. Therefore the condition 
for the quantized AHE is given only by the Hall
conductivity $\sigma_{xy}$ without the quantum correction, i.e., 
$|\sigma_{xy}| > e^2/(2h)$.
\end{abstract}
\pacs{72.15.Rn,73.43.-f,75.47.-m,75.70.-i}
\maketitle
The origin of the  anomalous Hall effect (AHE) has been a subject of
extensive controversy for a long term. One is based on
the band picture with the relativistic spin-orbit interaction
\cite{PR095_001154_54}, while the other
is due to the impurity scatterings \cite{PH024_000039_58}.
Most of the succeeding theories
follows the idea that the AHE occurs due to the scattering events modified
by the spin-orbit interaction, i.e., the skew scattering or the
side jump mechanism \cite{PTP27_000772_62}.
Recently several authors
recognized the topological nature of the AHE discussed in Refs.
\cite{PRL83_003737_99,JPSJ71_00019_02,PRL88_207208_02}.
In this formalism, the Hall conductivity $\sigma_{xy}$
is given by the Berry phase curvature in the momentum ($\vec k$-)
space integrated over the occupied states
\cite{PRL49_000405_82}.
Also there appeared some experimental evidences supporting it
\cite{JPSJ66_03893_97}.
Therefore it is very important to study the effect of
the scatterings by disorder, which makes $\vec k$ ill-defined, to see the
topological stability of this mechanism for AHE.

This issue is closely related to the integer quantum Hall effect (IQHE)
\cite{IQHE}
but there are several essential differences.
Usually the topological stability which guarantees the
quantization of some physical quantity, e.g., $\sigma_{xy}$, 
has been discussed in the context of the adiabatic continuation
\cite{IQHE}.
Therefore it appears that the gaps between Landau levels in pure system
are needed to start with 
even though the disorder potential eventually buries it.
In the IQHE system without disorder, the periodic potential is irrelevant 
because the carrier concentration is much smaller than unity per atom. 
Although numerical simulations
\cite{PRB64_165317_01} use lattice models, the main concern
is put on the limit of dispersionless
Landau levels separated by the gaps.
In the present case, i.e., in ferromagnetic metals,
there are multiple bands overlapping without the gaps in the density 
of states. The periodicity of the lattice remains unchanged,
which prohibits the uniform magnetic field and
also gives a large energy dispersion. In the language of the
effective magnetic field for electrons, it reaches a huge value
of the order of $\sim 10^4$ Tesla, i.e., the magnetic cyclotron length is
of the order of the lattice constant,
but the net flux is zero when averaged over the unit cell. 
Therefore these two cases belong to quite different limits
although the symmetries of the systems are common, i.e.,
the unitary class without time-reversal nor spin-rotational symmetry

In this paper we report on an extensive numerical study
on two-dimensional (2D) models of AHE including the disorder potentials.
It is found that the topological nature of AHE leads
to a dramatic phenomenon, i.e, the  ferromagnetic metal
turns into an integer quantized Hall system by introducing
disorder.  This is due to the topological stability
of the Chern numbers carried by the extended states which
are energetically separated by the continuum of the localized states
inbetween. Namely the localized state can not have a finite Chern number,
and the integer topological number can not change smoothly, i.e.,
it jumps when it changes. 
These two facts leads to the protection of the extended state
carrying a Chern number against the weak disorder.
The finite-size scaling analysis is compatible with
the two-parameter RG theory of Pruisken \cite{Pruisken},
which predicts the plateau transition at $|\sigma_{xy}| = 0.5e^2/h$. 
The critical exponents are consistent with that of the IQHE.
This problem is not an academic one;
the recent technology can fabricate very fine thin 
films of ferromagnetic metals with large enough single domain.
When the coherent length of such a thin film is 
longer than the thickness, it can be regarded as a multi-channel 2D system.
These systems can offer a possible laboratory to test our theory.

The essence of the AHE is that the Berry phase
of the Bloch electron is induced by the spin-orbit interaction
in the presence of the magnetization, which is modeled by the
complex transfer integrals \cite{JPSJ71_00019_02}.
Each band often gains finite Chern number
even though the density of states has no gap.
The minimal model describing this situation is that
proposed by Haldane \cite{PRL61_002015_88}
and its extension.
This model is defined on a honeycomb lattice containing two
atoms in a unit cell (Fig. 1).
Using the Fourier transformation
$\mbox{\boldmath{$c$}}^{\dagger}_{\vec{k}}
= (c^{\dagger}_{A \vec{k}}, c^{\dagger}_{B \vec{k}})$ of
the spinor representation
$(c^{\dagger}_{\vec{r}\in A}, c^{\dagger}_{\vec{r}\in B})$,
its Hamiltonian is written as
$H=\sum_{\vec{k}}
\mbox{\boldmath{$c$}}^{\dagger}_{\vec{k}}
{\cal H}_{\vec{k}}
\mbox{\boldmath{$c$}}_{\vec{k}}$, where
\begin{eqnarray}
{\cal H}_{\vec{k}}
&=& 2t_1\cos\phi\sum_{i=1,2,3} \cos(\vec{k}\cdot \vec{a}^{\rm
tr}_{i})\tau^{0}
\nonumber\\
&&
+t_0 \sum_{i}
\left[
\cos(\vec{k}\cdot\vec{a}^{\rm hc}_{i})\tau^{1}
-\sin(\vec{k}\cdot\vec{a}^{\rm hc}_{i})\tau^{2}
\right]
\nonumber\\
&&
+\left[
m+2t_1\sin\phi\sum_{i}\sin(\vec{k}\cdot\vec{a}^{\rm tr}_{i})
\right]\tau^{3}
\label{eq:Hamiltonian}.
\end{eqnarray}
$\tau^{0}$ is the unit matrix and $\tau^{1,2,3}$ are the Pauli matrices.
Here we assume the perfect spin-polarization
and use the spinless fermions.
The complex next-nearest neighbor hopping integral
$t_1 e^{i \phi} $ is introduced in addition to the real one $t_0$
between the nearest neighbors.
$\vec{a}^\mathrm{tr}_{1,2,3}$ are the lattice vectors of triangle sub-lattice,
while $\vec{a}^\mathrm{hc}_{1,2,3}$ are those of honeycomb lattice.

\begin{figure}
  \includegraphics[scale=0.7]{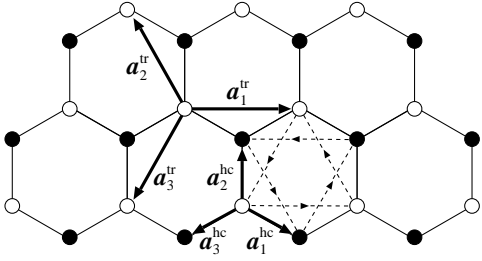}
  \caption{
Haldane's model defined on honeycomb lattice
\cite{PRL61_002015_88}.
Open and closed circles represent
the $A$ and $B$ sublattice sites respectively.
The thick arrows $\vec{a}^\mathrm{hc}_{1,2,3}$
and $\vec{a}^\mathrm{tr}_{1,2,3}$
are the lattice vectors of honeycomb lattice
and those of triangle sub-lattice respectively.
The dashed lines represent next-nearest-neighbor hopping.
}
\label{fig:model}
\end{figure}

\begin{figure}
  \includegraphics[scale=0.4]{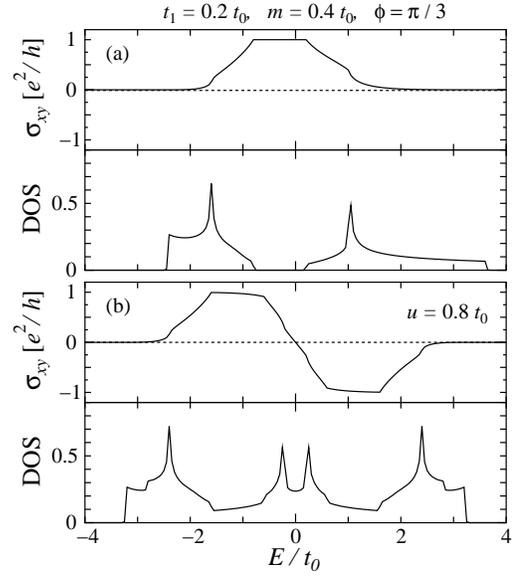}
  \caption{
Hall conductivity $\sigma_{xy}$ and
the density of states for the pure single-layer model with
$t_1 = 0.2t_0$, $m = 0.4t_0$ and $\phi = \pi/3$ (a),
and for the pure double-layer model with
$t_1 = 0.2t_0$, $m = 0.4t_0$, $u=0.8t_0$ and $\phi = \pi/3$ (b).
}
\label{fig:sigma-dos}
\end{figure}

The extended model is given by adding another layer
with the change $t_1\to -t_1$ to the original single-layer
model given above.
Furthermore we introduce the energy difference
between the layers by shifting
the uniform potential $\pm u$.
Then the extended model has the symmetric and gapless
density of states in contrast to the original one.
In Figs.~\ref{fig:sigma-dos} are shown the density of states and 
$\sigma_{xy}$ for the single-layer (a)
and double-layer (b) models, respectively.
In the single-layer case, $\sigma_{xy}$ is quantized to be
$e^2/h$ when the Fermi energy lies within the gap, while
it is not in the double-layer case where the
gap collapses.

\begin{figure}
    \includegraphics[scale=0.4]{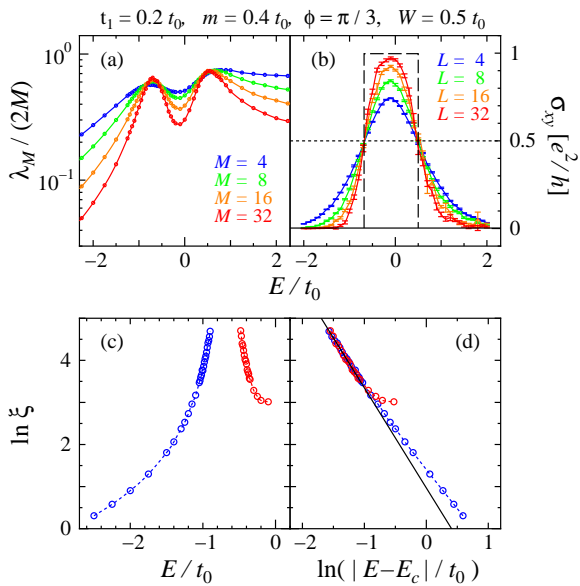}
  \caption{
(a) Localization length $\lambda_M$ of a quasi-1D tube
where $M$ is the number of A(B) sites on the circumference.
$M=4, 8, 16, 32$ are plotted.
(b) System-size dependence of $\sigma_{xy}$
in the single-layer system with $2L\times 2L$ lattice points.
The numbers of samples averaged are
$81920, 20480, 5120, 1280$ for $L= 4,8,16,32$
respectively.
The errors are one standard deviation.
(c) Log plot and (d) Log-log plot
of Localization length $\xi$
around the lower extended state at $E_c=(-0.69\pm 0.01)t_{0}$
The solid line is a fitting result with the slope $-\nu=-2.37$.
}
\label{fig:single-layer}
\end{figure}

Now we introduce the on-site disorder potential
to the single-layer model,
which is randomly distributed in the range $[-W/2,W/2]$,
and study the localization problem
in terms of the transfer matrix method \cite{ZPB53_0000001_83}.
Figure~\ref{fig:single-layer}(a) shows the dependence of the renormalized
localization length $\lambda_M/(2M)$ of a quasi-1D tube
with $2M$-sites circumference
on the strength of disorder potential $W$.
In each figure, the lines for $M=4, 8, 16, 32$ are plotted.
We have also calculated up to the system size  $M=64$
around the lower extended state.
(These additional data are used in the analyses
for the localization length and
its critical exponent.)
The length of a tube is typically $\sim 10^{5}$ sites
and the accuracy of data is within a few percent.
We can see that extended states are isolated in energy.
They merge with each other at
a critical value $6 < W_c < 7$,
and disappear, i.e., all states are localized.
This behavior is the same as that observed
in the ordinary IQH system on a square lattice
with external magnetic field \cite{PRB64_165317_01}.
It is noted that the pair annihilation of extended states
always occurs between those with the
opposite Chern numbers \cite{PRB64_165317_01}.
Actually, two extended states in Fig.~\ref{fig:single-layer}
originate from lower and upper bands
which have opposite Chern numbers, $\pm 1$, respectively.

Next we analyze the data 
to obtain a characteristic length $\xi(E,W)$,
which depends on $E$ and $W$ but not on $M$,
by the scaling hypothesis,
$
\lambda_M(E,W)/M = f\left(\xi(E,W)/M\right),
$
where $f(x)$ is a scaling function \cite{PRL42_000673_79}. 
As for a localized state,
$\xi(E,W)$ is interpreted as its localization length
in the thermodynamic limit.
Figure~\ref{fig:single-layer}(c) shows $\xi(E, W=5.0t_0)$ around
the lower extended state at $E=E_c$.
From this data, the critical exponent $\nu$ ($\xi \propto |E-E_c|^{-\nu}$)
is estimated as $\nu = 2.37 \pm 0.05$ with $E_c = (-0.69\pm 0.01)t_{0}$.
Figure~\ref{fig:single-layer}(d) gives
the log-log plot of the localization length $\xi(E, 5.0t_0)$
as a function of the energy measured from $E_c$.
The fitting is also shown as a solid line with
the slope $-\nu=-2.37$.
This value is in reasonable agreement with
that estimated in the ordinary IQH system \cite{RMP67_000357_95},
e.g. $\nu = 2.35 \pm 0.03$ \cite{EPL20_000451_92}.

The system-size dependence
of $\sigma_{xy}$ shown in
Fig.~\ref{fig:single-layer}(b)
represents its scaling property.
There are two critical points
where there is no size dependence and
which separate the two energy regions
with the opposite size dependences.
$\sigma_{xy}$ at these critical points
takes the value about $0.5e^2/h$.
This is consistent with the analysis in terms of the
effective field theory for the ordinary IQH system
in the weak-localization region \cite{Pruisken},
and strongly suggests that the critical properties of this transition
are same as those of the plateau transition
$\sigma_{xy} : 0\leftrightarrow 1$.
The energy of these critical points coincide
with that where the localization length diverges
in Figs.~\ref{fig:single-layer}(a) and (c).
This means that the extended states with Chern number $\pm 1$
exist there, and $\sigma_{xy}$ in the thermodynamic limit ($M \to \infty$)
stays quantized to be $e^2/h$ between these two energies [broken line
in Fig.~\ref{fig:single-layer}(b)].

\begin{figure}
  \includegraphics[scale=0.4]{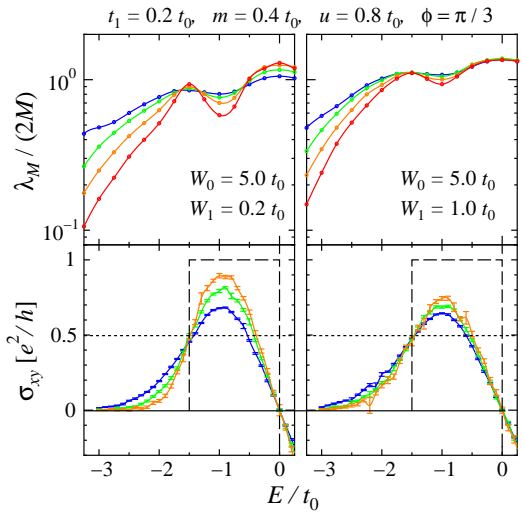}
  \caption{
Upper panels :
the localization length $\lambda_M$ of a quasi-1D tube
where $M$ is the number of A(B) sites on the circumference.
$M=4, 8, 16, 32$ are plotted.
Lower panels :
the system-size dependence of $\sigma_{xy}$
in the double-layer system with $2L\times 2L$ lattice points.
The numbers of samples averaged are
$81920, 20480, 5120$ for $L= 4,8,16$
respectively.
The color of each lines is for the same system-size
as in Fig.~\ref{fig:single-layer}(a) and (b).
}
\label{fig:double-layer}
\end{figure}

Now we shall consider the double-layer model.
The IQHE never occurs in this system with the above set of parameters
unless the effect of disorder is taken into account.
We consider the nontrivial case
in which there are scattering events both within and between these layers.
In this case, there is no gap between the initial and final
states of the elastic scattering, and it is
possible that all the states are localized once the
disorder is introduced. However as shown below, the
extended states and the Chern number carried by them are
stable against the weak disorder.
Here we define the strength of intra-layer scattering as $W_0$
and represent the strength of inter-layer scattering by $W_1$.
As seen in the upper panels of Fig.~\ref{fig:double-layer}, there occur
two energies, i.e., $E \cong - 1.5t_0$ and $E = 0$,
at which $\lambda_M/2M$ does not show $M$-dependence. This means
that the extended states survives there as in Figs. 3
at least up to $W_1 = 1.0 t_0$.

We next present the system-size dependence
of $\sigma_{xy}$, as shown in the lower panels
of Fig.~\ref{fig:double-layer}.
There appears two critical points for
the transitions $\sigma_{xy} : 0\leftrightarrow 1$
and $\sigma_{xy} : -1\leftrightarrow 1$.
From the particle-hole symmetry, $\sigma_{xy}(-E)=-\sigma_{xy}(E)$
is concluded.
Therefore, there appear three critical points
in the whole energy region.
Because the single-layer model has at most two critical-points,
the double-layer model could have maximum four.
One may wonder why there appear only three critical-points in the present
case.
This is because the middle one is composed of two extended states
which originally contribute to  different critical points
but carry the same Chern number, i.e. $-1$.
These extended states merge
(at least in the present numerical accuracy)
but never pair-annihilate, because the composite of
these extended states carry the non-zero Chern number $-2$.
In other words, the conservation law of the topological charge
prevents the localization.

$\sigma_{xy}$ at the lower critical-point
takes the value  $\cong 0.5e^2/h$.
This value is again consistent with
the analysis by Pruiskin and coworkers \cite{Pruisken}.
However, $\sigma_{xy}$ at the middle critical-point is zero.
This critical behavior seems
to violate the prediction by the analysis in \cite{Pruisken}.
However, recent numerical studies for the ordinary IQH system
reveals the new type of critical phenomena,
i.e. the direct transitions $\sigma_{xy} : 0\leftrightarrow n$ ($n>1$)
\cite{PRB64_165317_01},
which were experimentally observed in advance
\cite{PRL71_001439_93}.
The critical property around the middle point
is considered to belong to the same class as
$\sigma_{xy} : 0\leftrightarrow 2$.
Although the sample size is not large enough
in the double-layer model,
the size-dependence of $\sigma_{xy}$ is consistent with
the quantized plateau shown by the broken line
in the lower panels of Fig.~\ref{fig:double-layer}.

It is not difficult to generalize the
non-linear sigma model approach for the localization problem
to the case of multi-component model
without time-reversal nor spin-rotational symmetry.
This ``{\it components}'' means orbitals, spins, and channels
in the multilayer cases altogether.
This approach does not assume the finite gap at the starting.
Following the derivation in Ref.~\cite{Pruisken},
we obtain the Lagrangian:
\begin{eqnarray}
L[\{Q_l\}]
&=&
-\sum_{l,l'}\frac12[g^{-1}]_{ll'}\widetilde{\mathbf{Tr}}\;Q_{l}Q_{l'}
\nonumber\\
&&
+
\mathbf{Tr}\;\mathbf{ln}
\left[
E-\hat{H}+i\eta s
+ \sum_{l}i Q_l I_l
\right],
\label{eq:ef-model}
\end{eqnarray}
where $[g]_{ll'}$ is the scattering strength between 
components $l$ and $l'$, and $[I_l]_{l'l''}=\delta_{l'l}\delta_{l''l}$.
$\mathbf{Tr}$ ($\mathbf{ln}$)
is the trace (logarithm)
of matrix with functional index ($\vec{r}$)
and discrete indices ($p,a,l$),
where $p=\pm$ corresponds to the advanced and retarded fields respectively, 
and $a$ runs over replicas.
$\widetilde{\mathbf{Tr}}\;O$
is the abbreviation for
$\int d\vec{r}\;\mathrm{Tr}O(\vec{r})$
where $O(\vec{r})$ is a matrix with $p$ and $a$ indices.
The non-linear sigma model is the effective model
for the massless Goldstone modes.
In order to extract these modes, the parametrization
$Q_{l}=T_{l} P_{l} T^{-1}_{l}$ is useful.
From the above Lagrangian,
it is clear that inter-component scatterings $[g^{-1}]_{ll'}$ ($l\neq l'$)
lock the out-of-phase modes $T_{l}\neq T_{l'}$ ($l\neq l'$), 
and therefore the effective model for massless in-phase modes,
i.e., $T_{l} = T$,
reduces to the model identical to that
in Ref.~\cite{Pruisken}.
The coefficients of the stiffness
and topological terms for these modes
coincide with $\sigma_{xx}$ and $\sigma_{xy}$ respectively.
It is noted that these $\sigma_{xx}$ and $\sigma_{xy}$
contain the contributions from
all components, i.e., all orbitals, spins and channels.
Then the scaling of $\sigma_{xy}$ and $\sigma_{xx}$ remains
the same as given in Ref.~\cite{Pruisken}.
This supports the finite-size scaling study given above.

In real systems, the Coulomb interaction can not be neglected.
In IQHE system, the $\ln T$ dependence of $\sigma_{xx}$ is 
observed \cite{PRB25_005566_82} and is attributed to the quantum 
Coulomb correction \cite{PRB26_001651_82}.
However, the quantized $\sigma_{xy}$ in the ground state is 
well described by the noninteracting electron model.
The situation is similar here for the quantized AHE system. 
In the thin film of Fe, $\ln T$-dependence of $\sigma_{xx}$ is 
observed while not for $\sigma_{xy}$ \cite{PRL67_000735_91}, 
which is explained by the quantum Coulomb correction combined with the 
skew scattering mechanism  \cite{PRL67_000739_91}.

Usually AHE is estimated by 
$\rho_H \cong -\rho_{xx}^2 \sigma_{xy}$,
where $\rho_H$, $\rho_{xx}$ and $\sigma_{xy}$
are measured as quantities in 3D.
In good metals, $\rho_{xx}$ is very small at low temperatures, and hence
$|\sigma_{xy}|$ is large although $|\rho_H|$ is very small.
Therefore it is possible that the quantized AHE is realized even in
the conventional metallic ferromagnets such as Fe or Ni, when
the thin film is fabricated.
Actually, when we virtually consider thin film of $n$-layer systems, 
the 2D $|\sigma_{xy}|$ at $T_\mathrm{C}/2$ 
is estimated as $\sim 0.59 n e^2/h$ for Fe \cite{HPA25_000677_52},
$\sim 0.47 n e^2/h$ for Ni \cite{HPA25_000677_52},
and $\sim 0.20 n e^2/h$ for SrRuO$_3$
(from the first article in Ref.~\cite{JPSJ66_03893_97}).
Therefore, the condition $|\sigma_{xy}| > 0.5e^2/h$
is not so difficult to achieve in the thin films of metallic ferromagnets.
Extrapolating the $\ln T$ behavior of $\sigma_{xx}$ experimentally 
observed \cite{PRL67_000735_91}, 
the crossover temperature $T_\mathrm{cross}$ from weak to strong localization
is estimated as
$T_\mathrm{cross}\cong T_{0}
e^{-\frac{\sigma_{0}h}{Ae^2}}$,
where $T_{0}$ a reference temperature of the order of 10K, 
$\sigma_{0}$ the Drude conductivity at $T_{0}$,
and $A$ is a sample-dependent scaling-exponent of the order of unity. 
Therefore, if the minimal $\sigma_{0}h/(Ae^2)$ is less than $ \sim 10$,
we have the chance to observe the quantized anomalous Hall effect
in the experimentally realizable temperature. Considering that
$\sigma_{xy}$ have to be larger than $0.5e^2/h$, this condition means that
the ratio $\sigma_0/\sigma_{xy}$ should be smaller than $\sim 10$. 
This novel quantized Hall state would prove the most dramatic consequence 
of the topological nature of the AHE.

The authors would like to thank Y.~Tokura and A.~Asamitsu
for useful discussion.
M.~O. is supported by Domestic Research Fellowship
from Japan Society for the Promotion of Science.
N.~N. is supported by 
the Ministry of Education, Science, Culture and  Sports of Japan.



\end{document}